\documentclass[floatfix,aps,amsmath,nofootinbib,onecolumn,10pt]{revtex4}

\usepackage{graphicx}
\usepackage{bm}
\usepackage{rotating}
\usepackage{multirow}  
\usepackage{array}     
\usepackage{booktabs}  

\def\({\left(}
\def\){\right)}
\def\[{\left[}
\def\]{\right]}

\def\e{\begin{equation}}
\def\q{\end{equation}}
\def\m{\begin{eqnarray}}
\def\n{\end{eqnarray}}
\usepackage{hyperref}
\begin{document}

\title{Implications of the Two-Component Dark Energy Model for Hubble Tension }

\author{Lu Chen\footnote{Corresponding Author: chenlu@sdnu.edu.cn}, Peiyuan Xu, Guohao Li, Yang Han }
\affiliation{ School of Physics and Electronics, \\Shandong Normal University, Jinan 250014, China\\} 
\date{\today}

\begin{abstract}


Dark energy plays a crucial role in the evolution of cosmic expansion. In most studies, dark energy is considered a single dynamic component. In fact, multi-component dark energy models may theoretically explain the accelerated expansion of the universe as well. In our previous research, we constructed the $w_{\rm{n}}$CDM ($n=2, 3, 5$) models and conducted numerical research, finding strong observational support when the value of n is small. Based on our results, both the $\chi^2$ and Akaike information criterion (AIC) favor the $w_{\rm{2}}$CDM model more than the $w_0w_{\rm{a}}$CDM model. However, previous studies were limited to two equal-component dark energy models, failing to consider the component proportions as variables. Therefore, we will further explore the $w_{\rm{2}}$CDM model. To simplify the model, we fix $w = -1$ in one component and set the other component to $w_{\rm{de2}}$, varying the proportions of both components in the population. Each $w_{\rm{2}}$CDM model is constrained by the Planck 2018 TT, TE, EE + lowE + lensing, BAO data, and Pantheon dataset. Under different $w_{\rm{de2}}$, we obtain the one-dimensional distribution of ${H}_{0}$ with respect to $f_{\rm{de2}}$. Further fitting reveals the evolution of ${H}_{0}$ under varying $w_{\rm{de2}}$ and $f_{\rm{de2}}$. We perform the same operation on $\chi^2$, also obtaining the variation of $\chi^2$ with respect to $w_{\rm{de2}}$ and $f_{\rm{de2}}$. To evaluate the error of fitting, we introduce two indicators, $\text{R}^{2}_{\text{adj}}$ and MAPE, to quantify the fitting ability of our models. We find that when $w_{\rm{de2}}$ is less than -1, ${H}_{0}$ increases with the decrease of $w_{\rm{de2}}$ and the increase of $f_{\rm{de2}}$, effectively alleviating ${H}_{0}$ tension. For $\chi^2$, it still prefers the $\Lambda$CDM model, and the $w_{\rm{2}}$CDM model will decrease significantly when it approaches the $\Lambda$CDM model. The excellent performance of $\text{R}^{2}_{\text{adj}}$ and MAPE further proves that our model has an outstanding fitting effect and extremely high reliability.

\end{abstract}

\pacs{???}

\maketitle


\section{Introduction} 
\label{sec:int}

One of the most significant and fascinating discoveries in cosmology over the past three decades is that the universe is undergoing accelerated expansion \cite{SupernovaSearchTeam:1998fmf,SupernovaCosmologyProject:1998vns}. The underlying mechanisms behind cosmic acceleration have been a persistent challenge for the scientific community. Numerous cosmological hypotheses have been proposed to explain this phenomenon. Among these, the theory of dark energy (DE) has emerged as the most promising framework, gaining widespread acceptance among physicists. In general, DE can be characterized as constant, fluids, or fields \cite{Peebles:2002gy,Peebles:1987ek,Ratra:1987rm,Novosyadlyj:2010pg,Johnson:2020gzn,MohseniSadjadi:2020qnm,Kase:2020hst}, described by the dark energy equation of state (DE EOS) $w$. The standard model of cosmology, namely the $\Lambda$-Cold Dark Matter ($\Lambda$CDM) model \cite{Bamba:2012cp}, consists of the simplest DE model. In this model, DE is represented by the cosmological constant $\Lambda$, which corresponds to $w$ = -1. The $\Lambda$CDM model was extremely successful for a long time and brought many valuable insights to scientists, but it also encountered some serious theoretical difficulties \cite{Verde:2019ivm, DiValentino:2021izs, Riess:2021jrx, Wu:2024faw}. The most famous of these is the Hubble tension, which refers to the difference of 4.85$\sigma$ between ${H}_{0}$ = $67.36 \pm 0.54$ $\mathrm{km\cdot s^{-1}\cdot Mpc^{-1}}$ obtained by the cosmic microwave background (CMB) \cite{Planck:2018vyg}, and ${H}_{0}$ = $73.04 \pm 1.04$ $\mathrm{km\cdot s^{-1}\cdot Mpc^{-1}}$ from local measurements conducted by the SH0ES team \cite{Riess:2021jrx}. The $w$CDM model is a simple extension of the $\Lambda$CDM model, treating $w$ as a variable constant ($w\neq-1$). Other dynamic dark energy (DDE) models such as the ${w}_{0}{w}_{a}$CDM model \cite{Chevallier:2000qy,Linder:2002et}, the Barboza-Alcaniz (BA) model \cite{Barboza:2008rh} and the exponential (EXP) model \cite{Dimakis:2016mip,Pan:2019brc}, treat $w$ as a function of $z$. Notably, some DDE models have also received certain degrees of preference in specific combinations of observational data. \cite{DESI:2025zgx,DESI:2025qqy,DESI:2025fii,Giare:2024gpk,Arora:2025msq,Xu:2025nsn}.

However, most of the above models treat DE as a single component. We have already explored the possibility of multi-component DE models ($w_{\rm{n}}$CDM) in our previous work \cite{Wang:2020lpp}. According to our results, the $w_{\rm{n}}$CDM ($n=2, 3, 5$) models may outperform the DDE models when the number of free parameters is the same. Furthermore, analysis of $\chi^2$ and AIC in $w_{\rm{n}}$CDM models reveal that the two equal-component DE model performs slightly better than the $w_0w_{\rm{a}}$CDM model, although it did not surpass the base $\Lambda$CDM model and the $w$CDM model. However, previous research was limited to dividing DE components in equal proportions. Therefore, in the following, we undertake further research into the two-component DE model.
 
Our target is to alleviate the Hubble tension as much as possible and to expand the application scenarios of the $w_{\rm{2}}$CDM model. For simplicity, we conduct an in-depth study on the evolution of ${H}_{0}$ and $\chi^2$ in the $w_{\rm{2}}$CDM model with varying component ratios and different $w$. By comprehensively utilizing the CMB, BAO and SNIa data, as well as making targeted modifications to the CAMB program for different $w_{\rm{2}}$CDM models, we obtain accurate results for ${H}_{0}$ and $\chi^2$. Meanwhile, our analysis of the scatter plots for these two parameters shows distinct distribution patterns: $H_0$ exhibits obvious linear distribution characteristics, while $\chi^2$ follows a power function distribution. Through data fitting, we establish the functional relationships of ${H}_{0}$ and $\chi^2$ with respect to $w_{\rm{de2}}$ and $f_{\rm{de2}}$, respectively. In particular, the derived functional relationship of $H_0$ can significantly mitigate the Hubble tension under specific conditions. Simultaneously, we use two statistical indicators, $\text{R}^{2}_{\text{adj}}$ and MAPE, to analyze the fitting error, which shows that our model has high reliability.

This paper is organized as follows:
In section~\ref{sec:the}, we elaborate on the theoretical foundation of the $w_{\rm{2}}$CDM model.
In section~\ref{sec:res}, the datasets, methodology and our results are presented.
Finally, in section~\ref{sec:sum}, we summarize and discuss the main findings of this paper, and also illustrate the application prospects of the model.

\section{Theory}
\label{sec:the}

In this subsection, we extend the framework of multi-component DE models proposed in Ref.~\cite{Wang:2020lpp}.
In the $w_2$CDM model, the total density of DE today is artificially divided into 2 parts: i. one is assigned a constant equation of state $w_{\rm{de1}}=-1$ and accounts for $f_{\rm{de1}}$; ii. the other part is assigned a constant equation of state $w_{\rm{de2}}$ and accounts for $f_{\rm{de2}} =1-f_{\rm{de1}}$.

For background evolution of our universe, the total energy density is
\begin{equation}
\rho_{\rm{tot}}(a)
=\rho_{\rm{T}}(a) +\rho_{\rm{de}}(a)
=\rho_{\rm{T}} (a)+f_{\rm{de1}}\rho_{\rm{de}}(1)+f_{\rm{de2}}\rho_{\rm{de}}(1)a^{-3-3w_{\rm{de2}}}.
\end{equation}
Here, $\rho_{\rm{T}}$ is the energy density of all the other components of our universe, excluding DE. The second term indicates the evolution pattern of the first component of DE, and the third term represents that of the second component.

The parameterized post-Friedmann (PPF) \cite{Hu:2008zd,Fang:2008kc,Fang:2008sn,Grande:2008re,Li:2012spm} description is used to describe the evolution of the DE perturbation. Then the perturbation equations of the DE energy density and momentum are modified as follows,
\begin{eqnarray} 
\rho_{\rm{de}} \delta_{\rm{de}}&=&-3 \rho_{\rm{de}}^{\rm{w}} \dfrac{v_{\rm{de}}}{k_{\rm{H}}} - \dfrac{c_{\rm{K}} k_{\rm{H}}^2 H^2 \Gamma} {4\pi G},    \\
\rho_{\rm{de}} v_{\rm{de}}&=& \rho_{\rm{de}}^{\rm{w}} v_{\rm{T}} - \dfrac{k_{\rm{H}}^2 H^2} {4\pi G F}\[  S-\Gamma -\dfrac{\dot{\Gamma}}{H}\].
\end{eqnarray}
Here, we assume $\delta_{\rm{de}}\equiv \delta \rho_{\rm{de}}/\rho_{\rm{de}} =\delta_{\rm{de1}}\equiv \delta \rho_{\rm{de1}}/\rho_{\rm{de1}} =\delta_{\rm{de2}}\equiv \delta \rho_{\rm{de2}}/\rho_{\rm{de2}}$ for simplification.
Similarly, $v_{\rm{de}} =v_{\rm{de1}} =v_{\rm{de2}}$.
Besides, $G$ is Newton's constant and $H$ is the Hubble parameter. For a flat spatial universe, $c_{\rm{K}}=1$, $k_{\rm{H}}=k/aH$, where $k$ denotes the wave number in Fourier space. The symbol $\rho_{\rm{de}}^{\rm{w}}$ is defined as
\begin{equation}
\rho_{\rm{de}}^{\rm{w}}= f_{\rm{de2}}\rho_{\rm{de}}(1) a^{-3-3w_{\rm{de2}}} (1+w_{\rm{de2}}).
\end{equation}
Note that the term related to the first component of DE is gone because $1+w_{\rm{de1}}=0$. In addition,
\begin{eqnarray} 
F&=&1+ \dfrac{12\pi G a^2}{k^2 c_{\rm{K}}}  (\rho_{\rm{T}}+ p_{\rm{T}}),    \\
S&=& \dfrac{4\pi G}{H^2} \dfrac{v_{\rm{T}}+k\alpha}{k_{\rm{H}}}  \rho_{\rm{de}}^{\rm{w}} .
\end{eqnarray}
Here, $\alpha=a(\dot{h} +6 \dot{\eta})/2k^2$, where $h$ and $\eta$ represent the metric perturbations in synchronous gauge.
$\Gamma$ satisfies 
\begin{equation}
(1+c^2_{\Gamma}k^2_{\rm{H}}) \[ \Gamma +c^2_{\Gamma}k^2_{\rm{H}} \Gamma +\dfrac{\dot{\Gamma}}{H}  \] =S,
\end{equation}
where $c_{\Gamma}=0.4c_{\rm{s}}$ for scalar-field models and $c_{\rm{s}}$ denotes the sound speed.

\section{Datasets, Methodology and Results}
\label{sec:res}

Start at the beginning, 
we use the combined data of CMB, BAO and SNIa measurements for all our models in this work, which is given by the following:

\begin{itemize}
\item Measurements of CMB anisotropies from data released in 2018 by the Planck Collaboration, concretely, Planck 2018 plikHM$\_$TTTEEE$+$lowE$+$lensing~\cite{Planck:2018vyg,Planck:2018lbu}.
\item 
BAO measurements: 
6dFGS (the effective redshift $z_{\rm{eff}}=0.106$)~\cite{Beutler:2011hx},
MGS ($z_{\rm{eff}}=0.15$)~\cite{Ross:2014qpa}, the SDSS-III BOSS DR12 galaxy sample named DR12 ($z_{\rm{eff}}=0.38,0.51,0.61$)~\cite{BOSS:2016wmc}, the SDSS-IV eBOSS DR16 luminous red galaxy sample named DR16LRG ($z_{\rm{eff}}=0.70$)~\cite{eBOSS:2020hur}, the SDSS-IV eBOSS DR16 emission line galaxy sample named DR16ELG ($z_{\rm{eff}}=0.85$)~\cite{eBOSS:2020abk},  the SDSS-VI eBOSS DR16 quasar sample named DR16QSO ($z_{\rm{eff}}=1.48$)~\cite{eBOSS:2020uxp}, the SDSS-IV eBOSS DR16 auto-correlation named DR16Ly$\alpha$AUTO ($z_{\rm{eff}}=2.33$)~\cite{eBOSS:2020tmo}, as well as the cross-correlations of Ly$\alpha$ absorption and quasars named DR16Ly$\alpha$$\times$QSO ($z_{\rm{eff}}=2.33$)~\cite{eBOSS:2020tmo}.
\item The SNIa data, known as Pantheon, comprises 1048 samples and spans a redshift range of $z \sim 0.01-2.3$~\cite{Pan-STARRS1:2017jku}. 
\end{itemize}


We utilize the Monte Carlo Markov Chain (MCMC) nested sampler, along with the open-source packages CosmoMC and CAMB \cite{Lewis:1999bs,Lewis:2002ah,Lewis:2013hha}. 
Based on the analysis presented in Section II, we explore how $H_0$ changes under different $w_{\rm{de2}}$ and $f_{\rm{de2}}$. Specifically, we first set $w_{\rm{de2}}$ at six fixed values: \{$-0.9$, $-1.0$, $-1.1$, $-1.2$, $-1.3$, $-1.4$\}. Subsequently, for each predetermined value of $w_{\rm{de2}}$, we further adjust $f_{\rm{de2}}$ to \{$0$, $0.25$, $0.50$, $0.75$, $1$\} in sequence.
Obviously, when $f_{\rm{de2}} = 0$ or $w_{\rm{de2}} = -1$, all these $w_2$CDM models degenerate into the $\Lambda$CDM model. In contrast, when $f_{\rm{de2}} = 1$, the $w_2$CDM model is equivalent to the single-component $w$CDM model. For all cases except $w_{\rm{de2}} = -1$, we execute CosmoMC with six free parameters: \{${\Omega }_{\text{c}}{h}^{2}$, ${\Omega }_{\text{b}}{h}^{2}$, $\ln(10^{10})A_{\text{s}}$, ${n}_{\text{s}}$, $\tau$, $100{\theta }_{\text{MC}}$\} and set $\mathrm{action} = 2$ to obtain the best-fit values of $H_0$ and the corresponding $\chi^2$. To ensure the precision of constraints on the $\Lambda$CDM model, we run CosmoMC with $\mathrm{action} = 0$ to generate complete parameter chains, thus getting the marginalized errors of $H_0$ in the $\Lambda$CDM model. 

\subsubsection{${H}_{0}(w_{\rm{de2}}, f_{\rm{de2}})$ fitting}



\begin{figure}[]
\begin{center}
\includegraphics[scale=0.4]{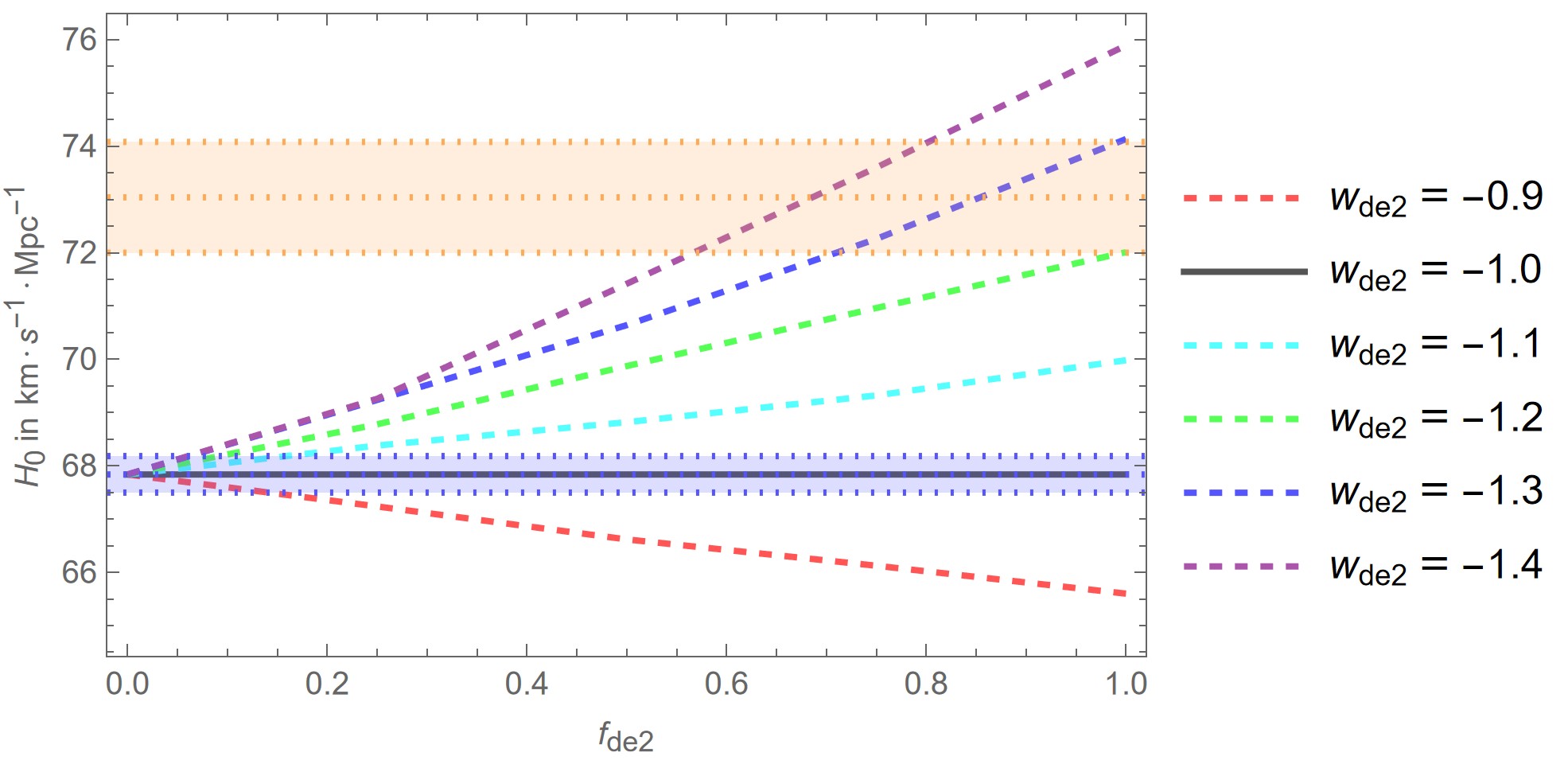}
\end{center}
\caption{The lines connecting the data points ($f_{\rm{de2}}$, ${H}_{0}$) under different values of $w_{\rm{de2}}$ are shown. Simultaneously mark the range of $H_0 = 73.04 \pm 1.04$ $\mathrm{km\cdot s^{-1}\cdot Mpc^{-1}}$ from SH0ES \cite{Riess:2021jrx} and $H_0 = 67.8365_{-0.3393}^{+0.3456}$ $\mathrm{km\cdot s^{-1}\cdot Mpc^{-1}}$ from our result.}
\label{fig1}
\end{figure}

With $f_{\rm{de2}}$ as an independent variable and $H_0$ as a dependent variable, we obtain five data points of the form $(f_{\rm{de2}}, H_0)$ for each given value of $w_{\rm{de2}}$. As shown in Figure~\ref{fig1}, we use the Mathematica program to draw lines connecting these data points and mark the range $H_0 = 73.04 \pm 1.04$ $\mathrm{km\cdot s^{-1}\cdot Mpc^{-1}}$ \cite{Riess:2021jrx}. The distribution of these data points approximates a straight line. Then a linear fitting is performed to derive an approximate linear functional relationship between $H_0$ and $f_{\rm{de2}}$:



\begin{equation}
H_0 \simeq 67.83646+k f_{\rm{de2}}\,[\,\mathrm{in\,km\cdot s^{-1}\cdot Mpc^{-1}\,}].  
\end{equation} Here, $k$ denotes the slope of each straight line. It is easy to see that when $w_{\rm{de2}} = -0.9$, $k$ is negative; when $w_{\rm{de2}} < -1$, $k$ becomes positive and increases as $w_{\rm{de2}}$ decreases.

To evaluate the error generated during the linear fitting, we introduce two indicators for error analysis. The first is the adjusted $\text{R}$-squared:
\begin{equation}
\text{R}^{2}_{\text{adj}}=1-\frac{\sum_{i=1}^{n} (y_{i}-\hat{y_{i}})^2 /(n-p-1) }{\sum_{i=1}^{n} (y_{i}-\bar{y} )^2/(n-1)}.  
\end{equation} Here, $n$ denotes the number of samples and $p$ represents the number of independent variables. Let $y_{i}$ be the true value, $\hat{y}_{i}$ the predicted value, and $\bar{y}$ the mean of the true values. $\text{R}^{2}_{\text{adj}}$ can be used to measure the proportion of data variation explained by the model. The closer its value is to 1, the better this model fits the data. The second indicator is the mean absolute percentage error ($\text{MAPE}$):
\begin{equation}
\text{MAPE}=\frac{1}{n} {\textstyle \sum_{i=1}^{n}}\left | \frac{y_{i}-\hat{y_{i} }}{y_{i}}  \right |\times100\%.
\end{equation} MAPE is a standardized metric that measures error as a percentage, directly reflecting the proportion of error to the true value. Generally, $\text{MAPE} < 10\%$ indicates excellent fitting performance and high model reliability. $10\% \le \text{MAPE} < 20\%$ corresponds good fitting performance, which can satisfy the requirements of most scenarios. $20\% \le \text{MAPE} < 50\%$ shows average fitting performance, requiring model optimization or data inspection. As for $\text{MAPE} \ge 50\%$, the fitting performance is poor, making the model unsuitable and requiring a redesign.

For different values of $w_{\rm{de2}}$, the specific results are given in Table~\ref{tab:fitting_results}. We find that both $\text{R}^{2}_{\text{adj}}$ and $\text{MAPE}$ yield excellent results, indicating that our fitting is highly successful.


\begin{table}[htbp]
    \centering
    \setlength{\tabcolsep}{8pt}
    \large
    \begin{tabular}{|l|l|c|c|c|}  
        \hline
        \multicolumn{3}{|c|}{\textbf{Fitting}} & \textbf{$\text{R}^{2}_{\text{adj}}$} & \textbf{MAPE} \\
        \hline
        \multirow{5}{*}{\textbf{Eq.(8)}} & $w_{\rm{de2}}$=-0.9 & $k$=-2.28 & 0.996 & 0.05\% \\
        \cline{2-5}
        & $w_{\rm{de2}}$=-1.1 & $k$=2.07 & 0.991 & 0.06\% \\
        \cline{2-5}
        & $w_{\rm{de2}}$=-1.2 & $k$=4.15 & 0.999 & 0.05\% \\
        \cline{2-5}
        & $w_{\rm{de2}}$=-1.3 & $k$=6.06 & 0.993 & 0.17\% \\
        \cline{2-5}
        & $w_{\rm{de2}}$=-1.4 & $k$=7.79 & 0.994 & 0.22\% \\
        \hline
        \multicolumn{3}{|c|}{\textbf{Eq.(11)}} & 0.997 & 3.83\% \\
        \hline
        \multicolumn{3}{|c|}{\textbf{Eq.(12)}} & 0.995 & 0.17\% \\
        \hline
    \end{tabular}
    \caption{The $k$ corresponding to different $w_{\rm{de2}}$ in Eq.(8), and the results of fitting error in Eq.(8), Eq.(11) and Eq.(12).}
    \label{tab:fitting_results}
\end{table}

\begin{figure}[]
\begin{center}
\includegraphics[scale=0.5]{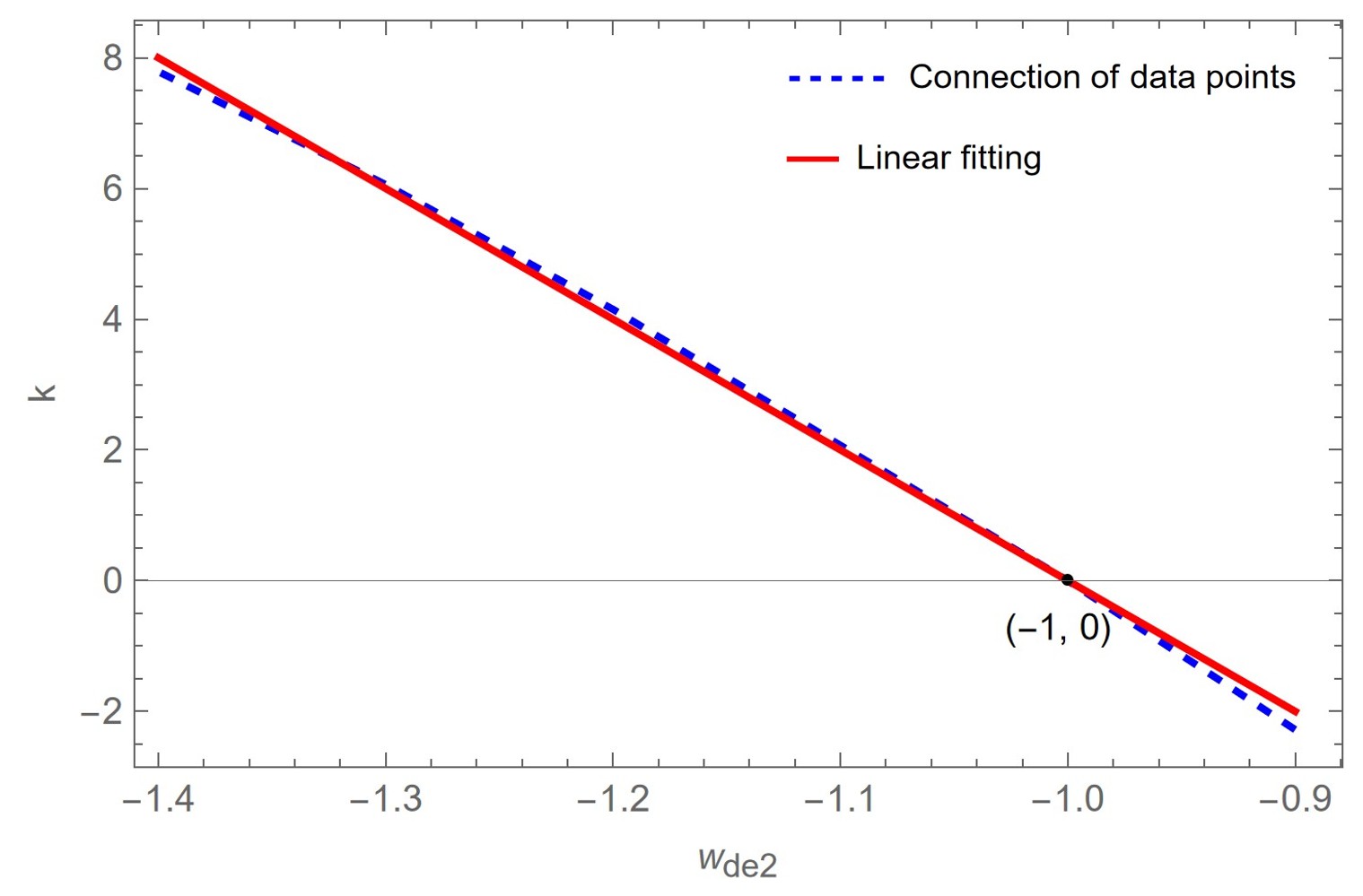}
\end{center}
\caption{The lines connecting the data points ($w_{\rm{de2}}$, $k$) and their corresponding linear fitting for Eq.(11).}
\label{fig2}
\end{figure}

Each value of $w_{\rm{de2}}$ corresponds to a specific $k$. Taking $w_{\rm{de2}}$ as the horizontal coordinate and $k$ as the vertical coordinate, we obtain six data points of the form $(w_{\rm{de2}}, k)$. As shown in Figure~\ref{fig2}, to ensure $k = 0$ when $w_{\rm{de2}} = -1$ (i.e., passing through the point $(-1, 0)$), we adopt the function $k = a + aw_{\rm{de2}}$ for the linear fitting. This yields the relationship between $k$ and $w_{\rm{de2}}$:


\begin{equation}
k \simeq -20.00 - 20.00 w_{\rm{de2}}.  
\end{equation}Under this linear fitting, $\text{R}^{2}_{\text{adj}} = 0.997$ and MAPE = 3.83\%, indicating an excellent fitting effect. 

Substituting Eq.(11) into Eq.(8), we can obtain ${H}_{0}$($w_{\rm{de2}}$, $f_{\rm{de2}}$), as follows:

\begin{equation}
H_0 \simeq 67.83646+(-20.00 - 20.00 w_{\rm{de2}}) f_{\rm{de2}}  =67.83646 -20.00 f_{\rm{de2}} -20.00 w_{\rm{de2}}f_{\rm{de2}} \,[\,\mathrm{in\,km\cdot s^{-1}\cdot Mpc^{-1}\,}].
\end{equation}

\begin{figure}[]
\begin{center}
\includegraphics[scale=0.6]{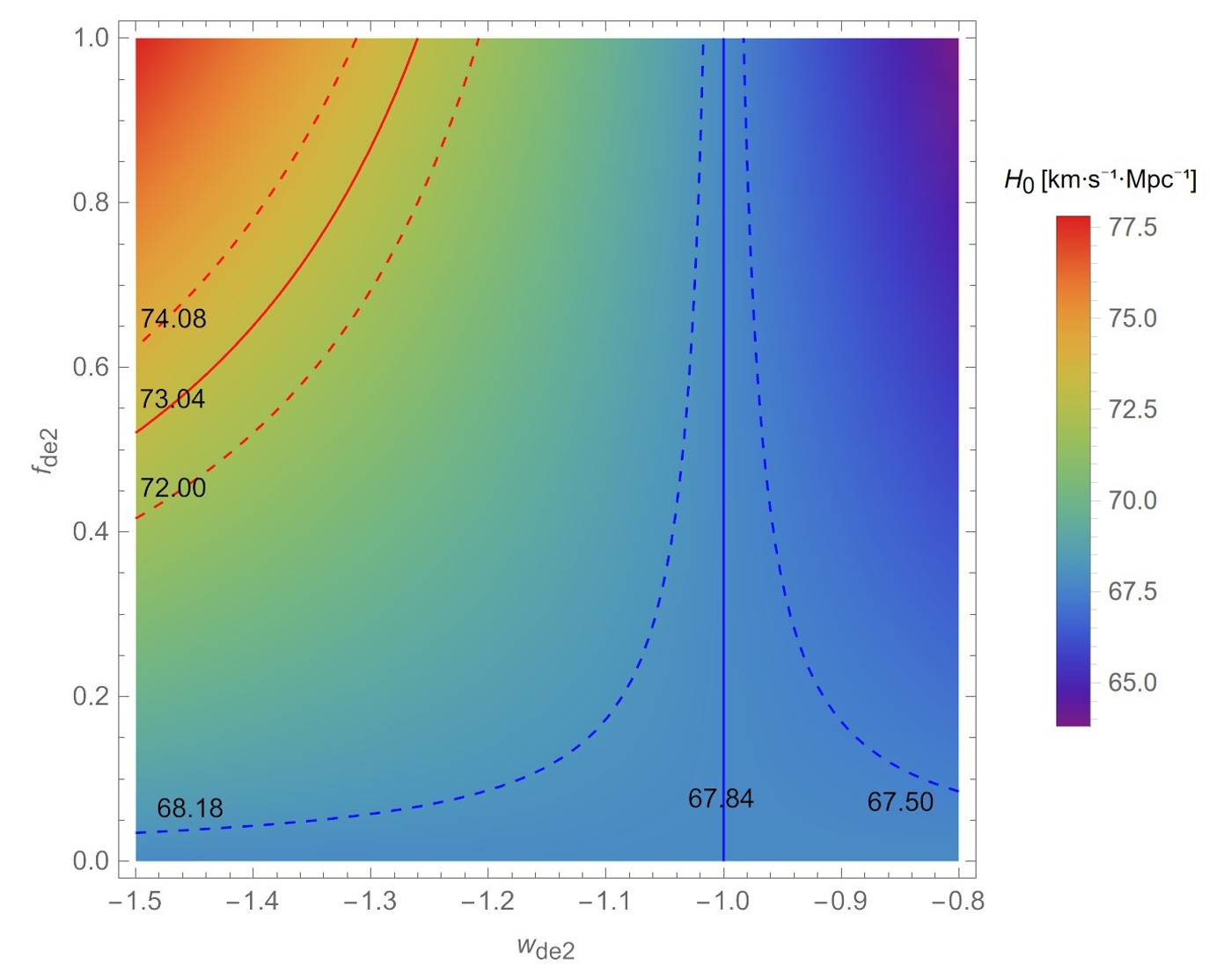}
\end{center}
\caption{The three-dimensional rainbow plot of ${H}_{0}$ with respect to $w_{\rm{de2}}$ and $f_{\rm{de2}}$. The ranges of $H_0 = 73.04 \pm 1.04$ $\mathrm{km\cdot s^{-1}\cdot Mpc^{-1}}$ from SH0ES \cite{Riess:2021jrx} and $H_0 = 67.8365_{-0.3393}^{+0.3456}$ $\mathrm{km\cdot s^{-1}\cdot Mpc^{-1}}$ from our result are also marked for reference.}
\label{fig3}
\end{figure}

As shown in Figure~\ref{fig3}, to more intuitively illustrate how $H_0$ varies with $w_{\rm{de2}}$ and $f_{\rm{de2}}$, we set $w_{\rm{de2}}$ as the horizontal axis and $f_{\rm{de2}}$ as the vertical axis. Gradient colors are used to represent $H_0$, where higher values are closer to red and lower values are closer to purple. We have also marked two ranges in this figure: $H_0 = 73.04 \pm 1.04$ $\mathrm{km\cdot s^{-1}\cdot Mpc^{-1}}$ \cite{Riess:2021jrx} and $H_0 = 67.8365_{-0.3393}^{+0.3456}$ $\mathrm{km\cdot s^{-1}\cdot Mpc^{-1}}$. The latter corresponds to the value in 68\% C.L. derived from CosmoMC in the $\Lambda$CDM model. The final error results for ${H}_{0}$($w_{\rm{de2}}$, $f_{\rm{de2}}$) are presented: $\text{R}^{2}_{\text{adj}}$ = 0.995, MAPE = 0.17\%. This shows that our model can explain 99.5\% of the variability in the dependent variable, exhibiting strong fitting ability. Meanwhile, the MAPE is significantly below the threshold of 10\%, indicating an excellent model fit.





\subsubsection{$\chi^{2}(w_{\rm{de2}},f_{\rm{de2}})$ fitting}

Next, we will perform the same operation on $\chi^{2}$ to find how it varies under different $w_{\rm{de2}}$ and $f_{\rm{de2}}$, thereby presenting the specific form of the fitted $\chi^{2}(w_{\rm{de2}},f_{\rm{de2}})$. The fitting process for $\chi^{2}$ is slightly complex. For simplicity, we only give the final total error of $\chi^{2}(w_{\rm{de2}},f_{\rm{de2}})$.

\begin{figure}[]
\begin{center}
\includegraphics[scale=0.4]{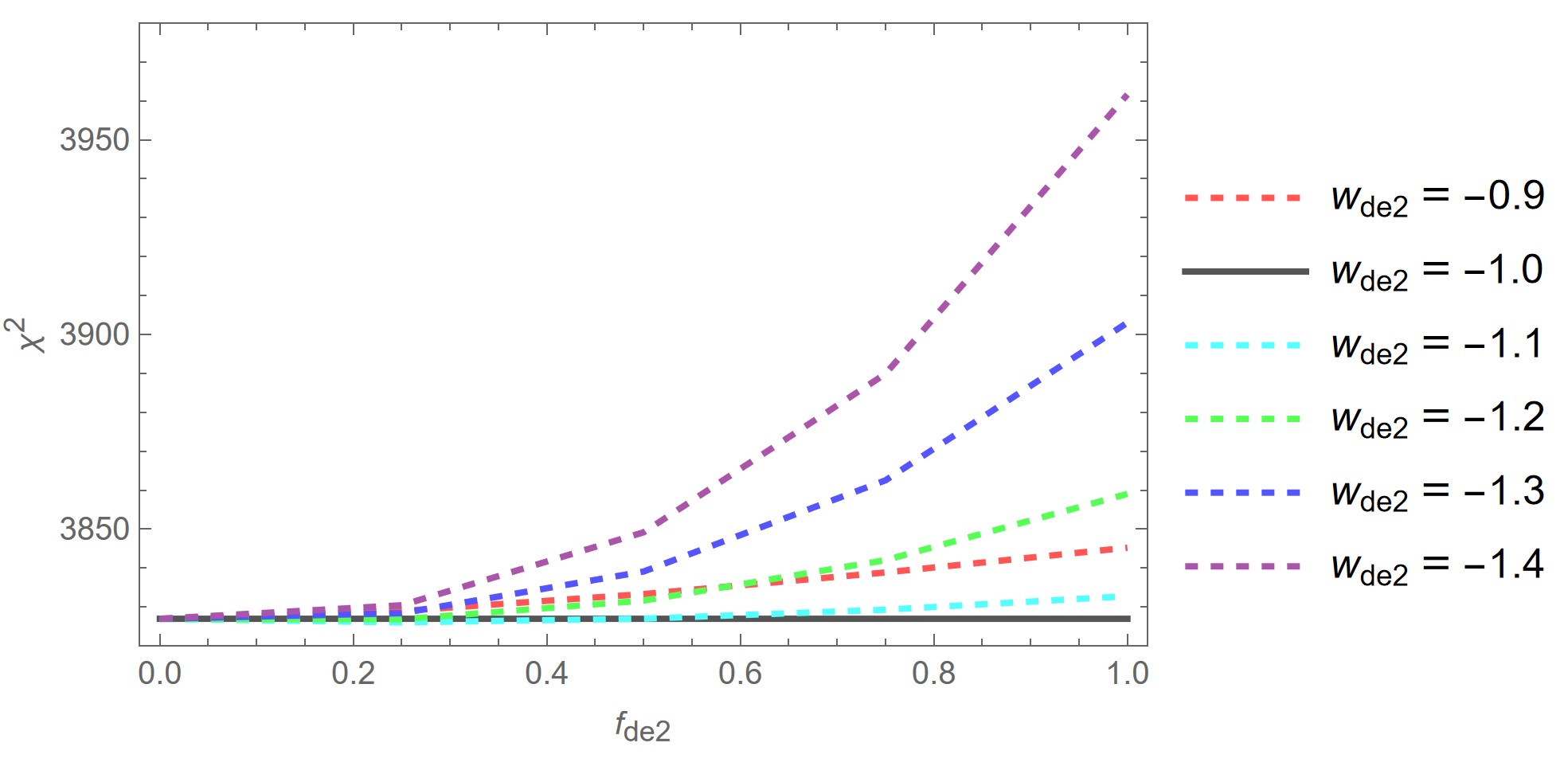}
\end{center}
\caption{The connecting lines of data points ($f_{\rm{de2}}$, $\chi^{2}$) under different $w_{\rm{de2}}$ conditions.}
\label{fig4}
\end{figure}

As shown in Figure~\ref{fig4}, with $f_{\rm{de2}}$ as the horizontal coordinate and $\chi^{2}$ as the vertical coordinate, we plot the lines that connect the data points with different values of $w_{\rm{de2}}$. Here, the data points exhibit an approximate quadratic function distribution. Therefore, through nonlinear fitting, we obtain the approximate quadratic function relationship between $\chi^{2}$ and $f_{\rm{de2}}$:

\begin{equation}
\chi^{2}\simeq{af_{de2}}^2+bf_{de2}+3826.9548.
\end{equation}
Here, we use two coefficients, $a$ and $b$, to describe this functional relationship. Similarly, for each value of $w_{\rm{de2}}$, there is a different set of $(a, b)$. 

\begin{figure}[]
\begin{center}
\includegraphics[scale=0.5]{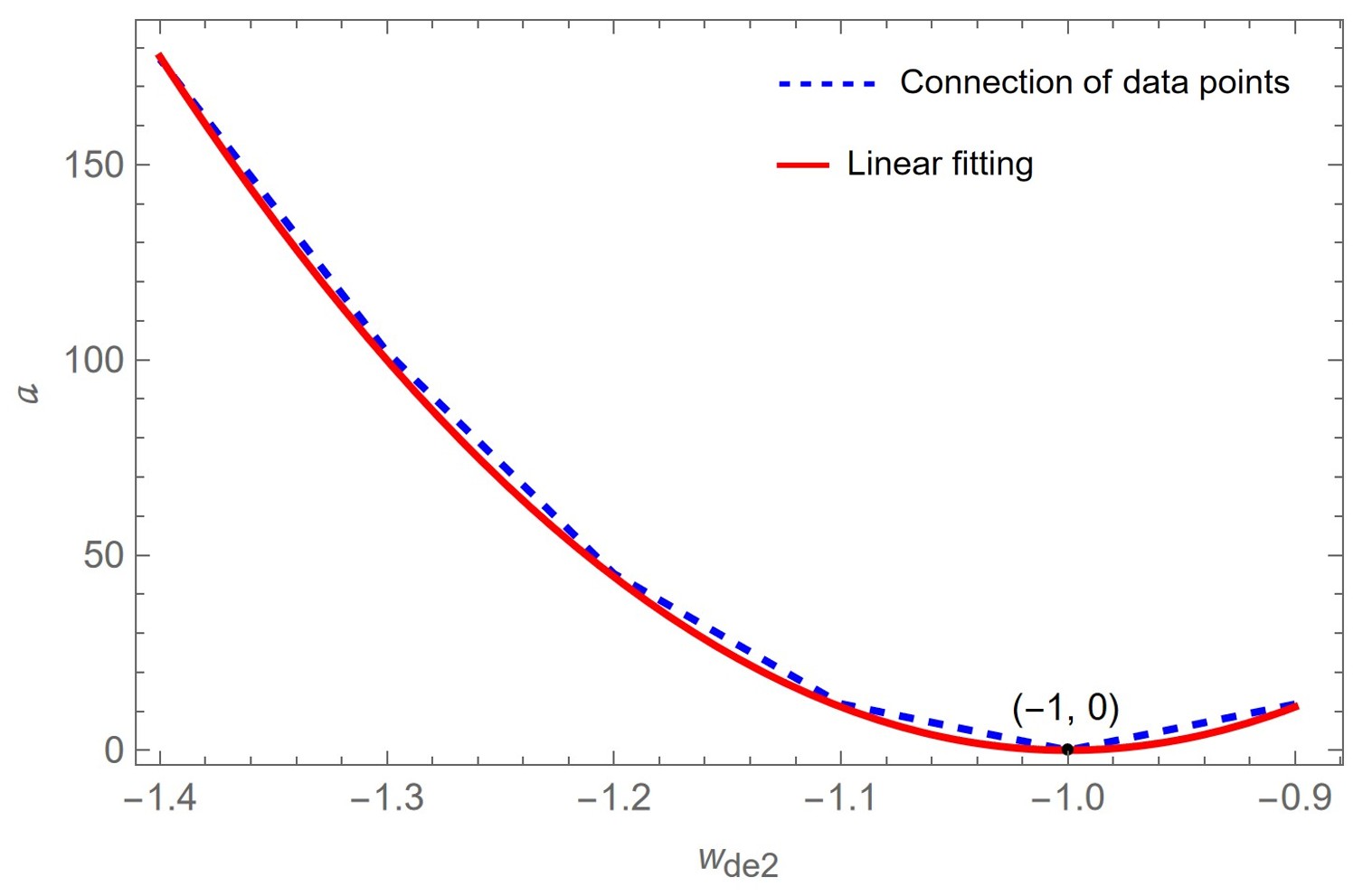}
\end{center}
\caption{The connection of data points ($w_{\rm{de2}}$, $a$) and the corresponding quadratic function fitting for Eq.(14).}
\label{fig5}
\end{figure}

Initially, we first analyze the relationship between the coefficient $a$ and $w_{\rm{de2}}$. As depicted in Figure~\ref{fig5}, the distribution of these data points approximates a quadratic function. To ensure that the data points pass through the point (-1, 0), we employ equation $a=\alpha(w_{de2}+1)(w_{de2}+\beta)$ for fitting. The result is as follows:

\begin{equation}
a\simeq1110.00(w_{de2}+1)(w_{de2}+1.00)=1110.00(w_{de2}+1)^2.
\end{equation}

\begin{figure}[]
\begin{center}
\includegraphics[scale=0.5]{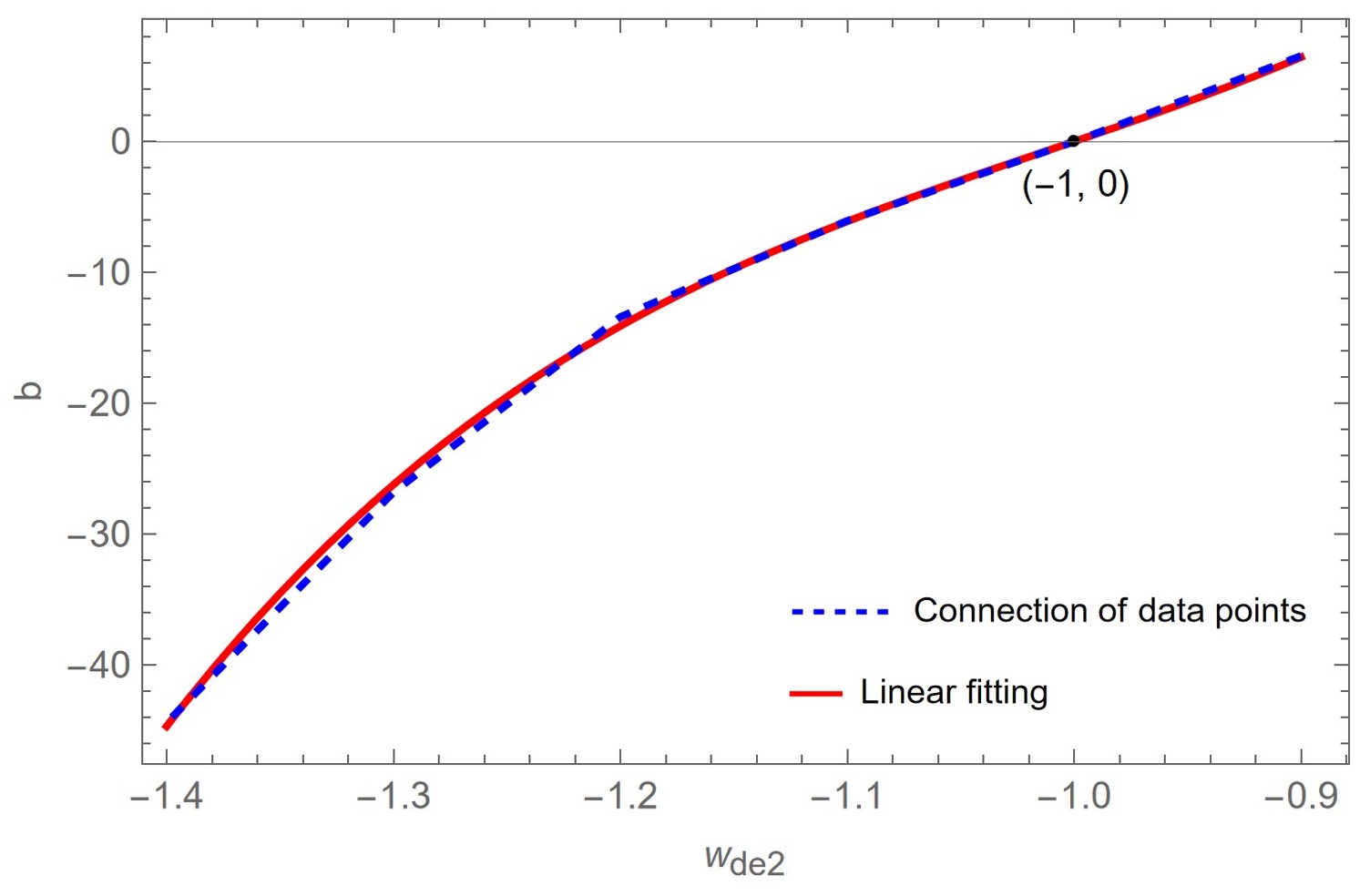}
\end{center}
\caption{The line connecting the data points ($w_{\rm{de2}}$, $b$) and the corresponding cubic function fitting for Eq.(15).}
\label{fig6}
\end{figure}

Then, we explore the variation of coefficient b with respect to $w_{\rm{de2}}$. As shown in Figure~\ref{fig6}, the distribution of the data points closely matches the cubic function. Similarly, to fit the data path through the point (-1, 0), we use $b=(w_{\rm{de2}}+1)(\alpha w_{\rm{de2}}^2+\beta w_{\rm{de2}}+\gamma)$ to obtain the following functional relationship:

\begin{equation}
b\simeq(w_{de2}+1)(372.00w_{de2}^2+761.00w_{de2}+448.00).
\end{equation}

By substituting Eq.(14) and Eq.(15) into Eq.(13), we obtain the final fitted $\chi^{2}(w_{\rm{de2}},f_{\rm{de2}})$ functional relationship:

\begin{equation}
\chi^2\simeq1110.00(w_{de2}+1)^2f_{de2}^2+(w_{de2}+1)(372.00w_{de2}^2+761.00w_{de2}+448.00)f_{de2}+3826.9548.
\end{equation}

\begin{figure}[]
\begin{center}
\includegraphics[scale=0.8]{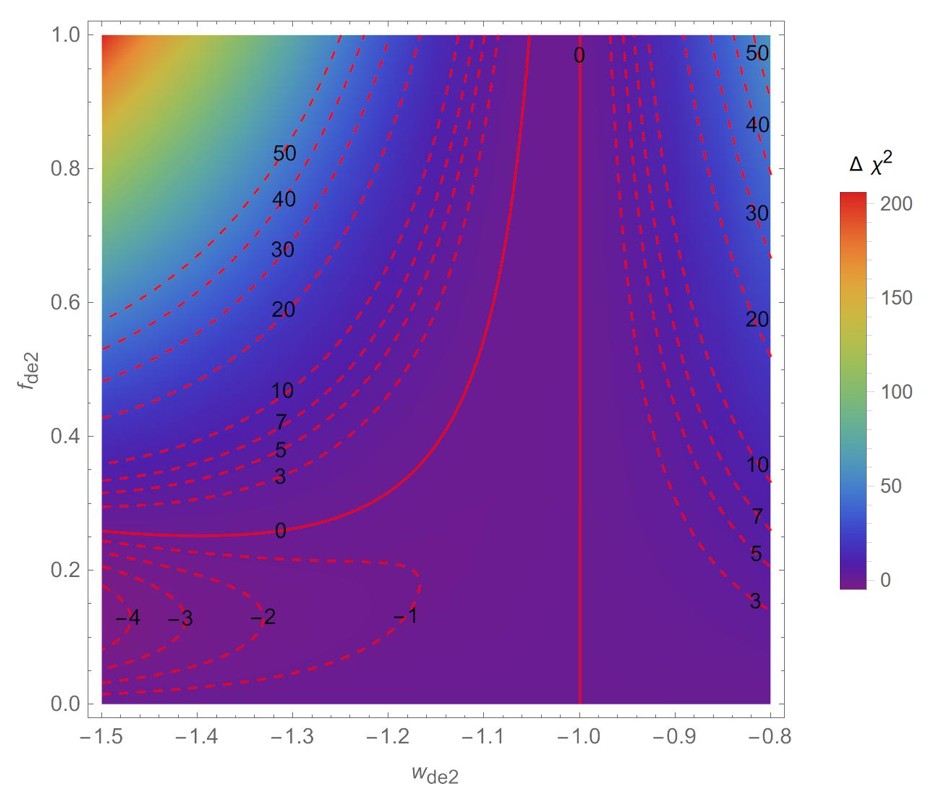}
\end{center}
\caption{The three-dimensional rainbow plot of $\Delta{\chi }^{2}$ with respect to $w_{\rm{de2}}$ and $f_{\rm{de2}}$. We have marked the contour lines of $\Delta{\chi }^{2}$ = -4, -3, -2, -1, 0, 3, 5, 7, 10, 20, 30, 40, and 50 in detail at this figure.}
\label{fig7}
\end{figure}

The difference between Eq.(16) and the $\chi^{2}$ of $\Lambda$CDM model is $\Delta{\chi }^{2}$ = $1110.00(w_{de2}+1)^2f_{de2}^2+(w_{de2}+1)( 372.00w_{de2}^2 +761.00w_{de2}+ 448.00) f_{de2}$. As shown in Figure~\ref{fig7}, we also use a rainbow graph to illustrate the variation of $\Delta{\chi }^{2}$ with respect to $w_{\rm{de2}}$ and $f_{\rm{de2}}$. Multiple contour lines with $\Delta{\chi }^{2}$ below 50 are marked in figure 7. Specifically, they are $\Delta{\chi }^{2}$ = -4, -3, -2, -1, 0, 3, 5, 7, 10, 20, 30, 40, 50. The trend is quite obvious, with a positive correlation between $\Delta{\chi }^{2}$ and $f_{\rm{de2}}$, and when $w_{\rm{de2}}$ approaches -1, $\Delta{\chi }^{2}$ decreases significantly. It can be seen that $\chi^{2}$ in the bottom left corner of this figure is slightly lower than that of the $\Lambda$CDM model, but the result is not significant.
Based on the final error results of $\chi^{2}(w_{\rm{de2}},f_{\rm{de2}})$, $\text{R}^{2}_{\text{adj}} = 0.998$, indicating that the model's ability to explain the variability of the dependent variable is as high as 99.8\%. The MAPE value of 0.02\% is also significantly lower than the threshold standard of 10\%, further confirming the excellent performance of model fitting.

\section{Summary and Discussion}
\label{sec:sum}

In this paper, we investigate a two-component DE model ($w_{\rm{2}}$CDM model), which incorporates two distinct DE components. The first component has a constant EOS with a value of $-1$ and accounts for $f_{\rm{de1}}$. The second component, with a constant EOS of $w_{\rm{de2}}$, accounts for $f_{\rm{de2}} = 1 - f_{\rm{de1}}$. The core of this study is to explore the variation patterns of ${H}_{0}$ and $\chi^{2}$ under different combinations of $w_{\rm{de2}}$ and $f_{\rm{de2}}$. Based on the modified open-source code CosmoMC + CAMB, combined with Planck 2018 TT, TE, EE + lowE + lensing, BAO and Pantheon datasets, we have completed the constraints of ${H}_{0}$ and $\chi^{2}$ under different $w_{\rm{2}}$CDM models. The results reveal clear regularities in the distributions of $H_0$ and $\chi^2$. Through data fitting, we establish the functional relationships between these quantities and the values of $w_{\rm{de2}}$ and $f_{\rm{de2}}$. Two statistical error metrics, $\text{R}^{2}_{\text{adj}}$ and MAPE, are introduced to quantify the fitting performance of the derived functions.

Our results indicate that at different values of $w_{\rm{de2}}$, ${H}_{0}$ and $f_{\rm{de2}}$ exhibit a clear one-dimensional linear distribution. Based on Figure~\ref{fig3} and the final fitting result of Eq.(12), it can be seen that when $w_{\rm{de2}} < -1$, a smaller $w_{\rm{de2}}$ combined with a larger $f_{\rm{de2}}$ leads to a higher $H_0$ value, which can effectively alleviate Hubble tension. In contrast, when $w_{\rm{de2}} > -1$, regardless of how $f_{\rm{de2}}$ and $w_{\rm{de2}}$ change, the value of ${H}_{0}$ is less than 67.83646 $\mathrm{km\cdot s^{-1}\cdot Mpc^{-1}}$, which further exacerbates the Hubble tension. With respect to the $\chi^{2}$ value, for a given $w_{\rm{de2}}$, $\chi^{2}$ exhibits a distribution characteristic similar to a quadratic function with $f_{\rm{de2}}$. The final fitting result of Eq.(16) shows that $\chi^{2}$ satisfies a power function relationship with $w_{\rm{de2}}$ and $f_{\rm{de2}}$. Compared to the $\Lambda$CDM model, all $w_{\rm{2}}$CDM models have no significant advantage in the $\chi^{2}$ values. When $w_{\rm{de2}}$ approaches -1, a smaller $f_{\rm{de2}}$ (i.e., the closer the model is to the $\Lambda$CDM model) results in a lower $\chi^2$ value, suggesting that the model is more favored in these cases.

In terms of data combination, we do not update the programs with the latest data~(such as Planck 2018 PR4 \cite{Rosenberg:2022sdy,Efstathiou:2019mdh,Carron:2022eyg}, DESI DR2 \cite{DESI:2025zgx,DESI:2025zpo}, DES-Y5\cite{DES:2024jxu,DES:2024upw,DES:2024hip}, et.al.). The reason is that our fitting process uses the best-fit value, and the variation of ${H}_{0}$ and $\chi^{2}$ with respect to $w_{\rm{de2}}$ and $f_{\rm{de2}}$ is determined by the $w_{\rm{2}}$CDM model. The new data can provide a smaller error bar, but the change in the best-fit value is not significant, leading to little impact on our conclusion. Additionally, for the early-stage exploration of this highly uncertain model, we consider that putting forward conservative conclusions may be more significant than proposing radical ones. 

The findings of this study have a clear application value and can be used directly in the $w_{\rm{2}}$CDM model. Specifically, given $w_{\rm{de2}}$ and $f_{\rm{de2}}$, our results can achieve a rapid numerical estimation of both ${H}_{0}$ and $\chi^{2}$, providing a practical tool for subsequent related studies. However, there are also areas where our work can be further improved. In terms of parameter values, $f_{\rm{de2}}$ only selected 5 special values and $w_{\rm{de2}}$ only set 6 fixed values. The relatively limited sample size may lead to a certain degree of randomness in the fitting results of $\chi^{2}$. Additionally, the choice of fitting function is not unique. If other function forms are used, the same ideal fitting effect may also be obtained, which needs further research in the future. Of course, we also consider this issue and use $\text{R}^{2}_{\text{adj}}$, an error evaluation indicator for small sample sizes, along with MAPE to precisely quantify the error magnitude. Ultimately, both metrics give convincing results, which improved the credibility of our model to some extent.

Finally, our work provides a highly accurate initial model, supporting the iteration of the $w_{\rm{2}}$CDM model and the updating of the cosmological observational data. In the future, if the sample size is appropriately increased and the value ranges of $w_{\rm{de2}}$ and $f_{\rm{de2}}$ are expanded, more accurate and convincing fitting results can be obtained.


\vspace{5mm}
\noindent {\bf Acknowledgments}

Lu Chen is supported by grants from NSFC (grant No. 12105164). This work has also received funding from project ZR2021QA021 supported by Shandong Provincial Natural Science Foundation and the Youth Innovation Team Plan of Colleges and Universities in Shandong Province (2023KJ350).



\end{document}